\documentstyle[12pt]{article}
 \def\appendix{\par
 \setcounter{section}{0}
 \setcounter{subsection}{0}
 \def\thesection{\Alph{section}}
 \def\theequation{\thesection.\arabic{equation}}}
 \def\thebibliography#1{\subsection*{References}\list
 {[\arabic{enumi}]}{\settowidth\labelwidth{[#1]}
 \leftmargin\labelwidth
 \advance\leftmargin\labelsep
 \usecounter{enumi}}
 \def\newblock{\hskip .11em plus .33em minus .07em}
 \sloppy\clubpenalty4000\widowpenalty4000
 \sfcode`\.=1000\relax}

\def\a{\alpha}
\def\b{\beta}
\def\d{\delta}
\def\e{\epsilon}

\def\h{\eta}
\def\k{\kappa}
\def\l{\lambda}
\def\m{\mu}

\def\o{\omega}

\def\q{\theta}
\def\r{\rho}
\def\s{\sigma}
\def\t{\tau}
\def\ups{\upsilon}

\def\z{\zeta}

\def\F{\Phi}
\def\G{\Gamma}

\def\Ld{\Lambda}
\def\O{\Omega}

\def\UPS{\Upsilon}
\def\X{\Xi}

\def\inbar{\vrule height1.5ex width.4pt depth0pt}
\def\rlx{\relax\leavevmode}
\def\I{\leavevmode\hbox{\small1\kern-3.8pt\normalsize1}}
\def\openone{\leavevmode\hbox{\small1\kern-3.3pt\normalsize1}}
\def\Ione{\rlx{\rm 1\kern-2.7pt l}}
\font\cmss=cmss10
\font\cmsss=cmss10 at 7pt
\def\ZZ{\rlx\leavevmode
             \ifmmode\mathchoice
                    {\hbox{\cmss Z\kern-.4em Z}}
                    {\hbox{\cmss Z\kern-.4em Z}}
                    {\lower.9pt\hbox{\cmsss Z\kern-.36em Z}}
                    {\lower1.2pt\hbox{\cmsss Z\kern-.36em Z}}
               \else{\cmss Z\kern-.4em Z}\fi}
\def\Ik{\rlx{\rm I\kern-.18em k}}  
\def\IC{\rlx\leavevmode
             \ifmmode\mathchoice
                    {\hbox{\kern.33em\inbar\kern-.3em{\rm C}}}
                    {\hbox{\kern.33em\inbar\kern-.3em{\rm C}}}
                    {\hbox{\kern.28em\sinbar\kern-.25em{\rm C}}}
                    {\hbox{\kern.25em\ssinbar\kern-.22em{\rm C}}}
             \else{\hbox{\kern.3em\inbar\kern-.3em{\rm C}}}\fi}
\def\IP{\rlx{\rm I\kern-.18em P}}
\def\IR{\rlx{\rm I\kern-.18em R}}
\def\IN{\rlx{\rm I\kern-.20em N}}

\def\llsymbol#1{\@llsymbol{\@nameuse{c@#1}}}
\def\@llsymbol#1{\ifcase#1\or {}\or {'}\or {''}\or {'''}\or
   {''''}\or {'''''}\or  \else\@ctrerr\fi\relax}

\newcounter{contador}

\def\acknowledgement{\if@twocolumn
\section*{Acknowledgements}
\else \normalsize
\begin{center}
{\bf Acknowledgements\vspace{-.5em}\vspace{0pt}}
\end{center}
\quotation
\fi}
\def\endacknowledgement{\if@twocolumn\else\endquotation\fi}

\newcommand{\ol}\overline
\newcommand{\ti}\tilde
\newcommand{\wt}\widetilde
\newcommand{\wh}\widehat
\newcommand{\bv}\breve
\newcommand{\dg}\dagger

\newcommand{\be}{\begin{equation}}
\newcommand{\ee}{\end{equation}}
\newcommand{\bl}{\begin{eqnarray}&}
\newcommand{\el}{&\end{eqnarray}}
\newcommand{\bq}{\begin{eqnarray}}
\newcommand{\eq}{\end{eqnarray}}
\newcommand{\0}{{\bf 0}}

\newcommand{\ad}{{\dot\alpha}}
\newcommand{\bd}{{\dot\beta}}

\newcommand{\uptad}{\widetilde\theta^{\dot\alpha}}

\newcommand{\qwt}{\widetilde\theta}
\newcommand{\ov}{\overline}
\newcommand{\pa}{\partial}

\begin{document}
\setcounter{page}{0}
\begin{center}
{\huge Gauging N=2 Supersymmetric Non-Linear $\boldmath{\s}$-Models in 
 the Atiyah-Ward Space-Time}
\vspace{15mm}

{\Large M. Carvalho\footnote{e-mail address:
kitty@cbpfsu1.cat.cbpf.br}}
\vspace{10mm}

{\Large and} 
\vspace{10mm}

{\Large M.W. de Oliveira\footnote{e-mail address:
 mwerneck@cbpfsu1.cat.cbpf.br}}
\vspace{15mm}

{\large Centro Brasileiro de Pesquisas F\'\i sicas (CBPF) \\
        Departamento de Campos e Part\'\i culas (DCP)\\
        Rua Dr. Xavier Sigaud, 150 - Urca \\
        22290-180 - Rio de Janeiro - R.J. - Brazil.}
\vspace{15mm}
 
CBPF--NF--073/96
\end{center}
\vspace{15mm}

\begin{center}
{\large{\bf Abstract}}
\end{center}

{We build up a class of N=2 supersymmetric non-linear $\s$-models 
in an N=1 superspace based on the Atiyah-Ward space-time of 
(2+2)--signature metric. We also discuss the gauging of isometries 
of the associated hyper-K\"ahlerian target spaces and present the 
resulting gauge-covariant supersymmetric action functional.}

\newpage 
\section{Introduction}

In the recent years much attention has been paid to the construction
of new classical field models in the Atiyah-Ward space-time of 
(2+2)--signature metric \cite{ww}. As demonstrated in refs.\cite{mm,ov}, 
this structure emerges in connection with a consistent N=2 superstring 
theory, whose underlying superconformal algebra requires a complex 
manifold as the relevant space-time background.

From the viewpoint of mathematics, the Atiyah-Ward space-time is also
quite attractive when regarded as a four-dimensional arena in which
one could introduce self-dual Yang-Mills connections \cite{wa,aw}. 
In fact, these objects are known to play a significant role as a 
field-theoretical tool in the Donaldson's programme on algebraic 
geometry \cite{don} and, as conjectured by Ward \cite{wb}, may be 
also of importance in the classification of lower-dimensional 
integrable models. 

In view of these facts, it seems also interesting to build up and
analyze supersymmetric Yang-Mills theories in the Atiyah-Ward 
space-time. Indeed, such models where first considered by Gates et al.
in refs.\cite{gkn}, where a superspace formalism adapted to the 
(2+2)-signature was introduced: the so-called N=1 superspace of 
Atiyah-Ward. Other related aspects in this domain were further 
investigated in ref.\cite{ac}. Moreover, in ref.\cite{chv}, one was 
able to present a supersymmetric non-linear $\s$-model also in the
Atiyah-Ward superspace and to couple its associated scalar superfields
to a super-Yang-Mills gauge sector through the gauging of isometries of 
the target manifold \cite{bw,bag,st,hklr,hiklr,bgio,ahs,hs,hps}. 
Clearly, the class of theories foccused here should be necessarily 
understood in the sense of the dimensional reduction framework used 
by Ward in \cite{wb}. In that scheme, one may eventually obtain new 
examples of integrable field models in two dimensions (see also 
ref.\cite{nis}). 

This is the purpose of the present work: to give a detailed account 
on the construction and gauging of supersymmetric $\s$-models \`a la 
Atiyah-Ward. Specifically, we will be concerned here with 
hyper-K\"ahlerian $\s$-models possessing N=2 supersymmetries 
-- one of them being non-linerarly realized -- and, subsequently, 
with the issue of performing their gauging by means of the approach 
developed in ref.\cite{hklr}.

Our paper is organized as follows: in Section 2 we describe in a 
self-contained fashion all the necessary steps needed to build up the 
gauged N=1 supersymmetric $\s$-model in D=2+2 dimensions (a problem 
already addressed in ref.\cite{chv}) and state the essential notions 
on hyper-K\"ahler geometry which are crucial for the N=2 extension of 
the following section; Section 3 is then devoted to the study of N=2  
supersymmetry in the N=1 superspace of Atiyah-Ward and to the gauging 
of the hyper-K\"ahlerian $\s$-model in the context of a certain 
K\"ahlerian vector supermultiplet. In Section 4 we interpret our 
results and present our conclusions.

\section{The hyper-K\"ahlerian $\boldmath{\s}$-Model in Superspace}

We begin the present investigation by foccusing on the construction 
of gauged N=1 supersymmetric $\s$-models in the Atiyah-Ward space-time. 
The notation and conventions for a superspace with base space-time 
possessing a $(2+2)$--signature are the same as in \cite{ac}. To build 
up the action functional for a class of K\"ahlerian $\s$-models one will 
follow here the well-known method of Zumino \cite{zum} (see 
refs.\cite{yan,hilp} for an extensive discussion on K\"ahler geometry).
We introduce a set of complex chiral and antichiral superfields, $\F^{i}$ 
and $\X^{i}$ ($i$=1,...,n), with their component field expansions written 
as{\footnote{The Grassmann coordinates, $\q\, {\mbox{and}}\, \tilde\q$,
are Majorana-Weyl spinors.}:
\be
\F^{i}= A^{i} + i\q\psi^{i} + i\q^{2} F^{i} 
        + i\tilde\q \tilde{\s}^{\m}\q \pa_{\m}A^{i}
        + \mbox{$\frac{1}{2}$}~
          \q^{2}\tilde\q \tilde{\s^{\m}}\pa_{\m}\psi^{i}
        - \mbox{$\frac{1}{4}$}~
          \q^{2}\tilde\q^{2}\Box A^{i},
\ee
\be
\X^{i}= B^{i} + i\tilde\q\tilde\chi^{i} + i\tilde\q^{2}G^{i} 
        + i\q {\s}^{\m}\tilde\q \pa_{\m} B^{i}
        +\mbox{$\frac{1}{2}$}~\tilde\q^{2}\q\s^{\m} {\pa}_{\m}\tilde\chi^{i}                                                                                         
        -\mbox{$\frac{1}{4}$}~\q^{2}\tilde\q^{2}\Box B^{i},
\ee
where $A^{i}$ and $B^{i}$ are complex scalar fields, $\psi^{i}$ and 
${\tilde{\chi}}^{i}$ are Majorana-Weyl spinors and $F^{i}$ and $G^{i}$ 
are complex scalar auxiliary fields. One has to observe that, differently 
to the Minkowskian situation, the scalar superfields at hand do not change 
their chirality properties under the complex conjugation operation: 
\bq
&{\wt{D}_{\ad}}\F^{i}={\wt{D}_{\ad}}\F^{*i}=0, \nonumber \\
&D_{\a}\X^{i}=D_{\a}\X^{*i}=0,
\label{quiral}
\eq
with
\bq
D_{\a}&=&\pa_{\a}-i\uptad {\pa}_{\a\ad},
\nonumber \\
{\wt{D}_{\ad}}&=&\wt{\pa}_{\ad}-i\q^{\a}{\wt{\pa}}_{\ad \a},
\label{derivatives}
\eq
and
\bq
&\{D_{\a},{\wt{D}_{\ad}}\}=-2i\;{\s}^{\m}_{\a
\ad}~{\pa}_{\m},~~~\{D_{\a},{{D_{\b}}}\}=\{{\wt{D}_{\ad}},
{\wt{D}_{\bd}}\}=0,
\nonumber \\
&[D_{\a},{\pa}_{\m}]=[{\wt{D}_{\ad}},{\pa}_{\m}]=0.
\label{dalgebra}
\nonumber
\eq
Now, one writes down a rather specific supersymmetric action to govern 
the dynamics of the scalar superfields. 
We take{\footnote{$\int{d^4xd^2{\q}d^2\qwt} 
\equiv \frac{1}{16}\int{d^4x}D^{\a}
 {\wt{D}^{\ad}}\wt{D}_{\ad}D_{\a}$}}:
\be
I=2\int{d^4x~d^2{\q}~d^2\qwt}~K(\F^{i},\X^{i};\F^{*i},\X^{*i}),
\label{action}
\ee
where the K\"ahler potential $K$ decomposes into two conjugated 
pieces as below:
\be
K(\F^{i},\X^{i};\F^{*i},\X^{*i})=H(\F^{i},\X^{*i})+H^*(\F^{*i},\X^{i}).
\label{deco} 
\ee
The pure scalar sector steming from the projection of (\ref{action}) 
into component fields is given by: 
\be
I_{scalar} =2 \int{d^{4}x}~ 
              \biggl(\frac{\pa^2 K}{\pa A^{i}\pa {B^{*j}}} 
              \pa_{\m}A^{i}\pa^{\m}{B^{*j}}
             +\frac{\pa^2 K}{\pa {A^{*i}}\pa B^{j}}\pa_{\m}{A^{*i}}
              \pa^{\m}B^{j} \biggr).
\label{action1}
\ee
Upon dimensional reduction and proper field truncations, $I_{scalar}$
above will give rise to a sensible (ghost free) scalar kinetic term in 
D=1+2 space-time dimensions (see ref.\cite{ac}).

The possible target spaces associated to the action $I$ in 
(\ref{action}) do belong to a restricted class of 4n-dimensional K\"ahler 
manifolds, their Hermitian metric tensor appearing in a four-block 
structure as follows:
\be
g_{{\cal I}{\cal J}} =
\left(
\begin{array}{cccc}
\0 &\0 &\0 & g_{i  \ov{\hat{\jmath}}} \\
\0 &\0 & g_{\hat{\imath} \ov{\jmath}} & \0 \\
\0 & g_{\ov{\imath} \hat{\jmath}} & \0& \0 \\
g_{\ov{\hat{\imath}} j}& \0& \0& \0 \\
\end{array}\right),
\label{metrica} 
\ee
with
\be
g_{i\ov{\hat{\jmath}}}=\frac{\pa^2 H}{\pa\F^{i}\pa {\X^{*j}}},
~~~~~
g_{\hat{\imath}\ov{\jmath}}=\frac{\pa^2 H^{*}}{\pa\X^{i}\pa {\F^{*j}}},
~~~~~
g_{\ov{\imath}{\hat{\jmath}}}=\frac{\pa^2 H^{*}}{\pa\F^{*i}\pa{\X^{j}}},
~~~~~
g_{\ov{\hat{\imath}}j}=\frac{\pa^2 H}{\pa\X^{*i}\pa{\F^{j}}},
\nonumber
\ee
and
\bq
{\cal I},{\cal J}=1,...4n~{\mbox{and}}~i,j=1,...n.   
\nonumber
\label{metrica2}
\eq
It is clear now that the particular form of $g_{{\cal I}{\cal J}}$ 
will entail a number of consequences for the geometry of our K\"ahlerian 
target manifold. The most general type of K\"ahler transformation one can
perform upon the potential $K$ while keeping the action (\ref{action})
invariant and the metric (\ref{metrica}) unchanged is:
\be
K \longrightarrow K^{'}=K +\h(\F) +\h^{*}(\F^*) +\r(\X) +\r^{*}(\X^*),
\label{trans1}
\ee
with $(\h,\h^*)$ and $(\r,\r^*)$ standing for arbitrary chiral and 
antichiral functions respectively. Hence, every isometry transformation 
of the target manifold will be a simmetry of (\ref{action}) provided its 
action on $K$ writes into a form compatible with (\ref{trans1}). The 
Killing vectors $({\k}^i_a (\F),{\t}^{i}_a (\X),{\k}^{*i}_a (\F^*),
{\t}^{*i}_a (\X))$ are the generators of the isometry group $\cal G$ 
and satisfy the usual Lie algebraic relations:
\bq
&{\k}^{i}_{a} {\k}^{j}_{b,i}-{\k}^{i}_{b} {\k}^{j}_{a,i}=                                          
 f_{ab}\,^{c}\,{\k}^{j}_{c},~~~~~
{\k}^{*i}_{a} {\k}^{*j}_{b,i}-{\k}^{*i}_{b} {\k}^{*j}_{a,i}=
f_{ab}\,^{c}\,{\k}^{*j}_c,& \nonumber \\
&{\t}^{i}_{a} {\t}^{j}_{b,i}-{\t}^{i}_{b} {\t}^{j}_{a,i}= 
f_{ab}\,^{c}\,{\t}^{j}_{c},~~~~~
{\t}^{*i}_{a} {\t}^{*j}_{b,i}-{\t}^{*i}_{b} {\t}^{*j}_{a,i}= 
f_{ab}\,^{c}\,{\t}^{*j}_{c},&
\label{lie}
\eq
where $f_{ab}\,^{c}$ are the structure constants. A global
isometry transforms the target coordinates as:
\bq
&\F^{'i}  =  \exp{(L_{\l \cdot \k})} \F^{i},~~~~~
\F^{'*i}  =  \exp{(L_{\l \cdot \k^{*}})} \F^{*i},& \nonumber \\
&\X^{'i}  =  \exp{(L_{\l \cdot \t})} \X^{i},~~~~~
\X^{'*i}  =  \exp{(L_{\l \cdot \t^{*}})} \X^{*i},&
\label{global}
\eq
where $\l$ is for a real parameter and $L_{\l.\k}$ (resp. $L_{\l.\t}$) 
is the Lie derivative along the vector field 
$\l.\k\equiv \l^{a}\k^{i}_{a}\pa_{i}$ (resp.
$\l.\t\equiv \l^{a}\t^{i}_{a}\pa_{\hat{\imath}}$). 
The set of laws above may be related to some, K\"ahler 
transformation like (\ref{trans1}), the chiral and antichiral 
functions being given as:
\bq
\h_a (\F) & = &\pa_{i}H(\F,\X^*)~{\k}^i_a (\F) +Y_a(\F,\X^*),\nonumber\\
\r_a (\X) & = & \pa_{\hat{\imath}}{H^*}(\X,\F^*)~{\t}^i_a (\X) 
- Y^*_a (\X,\F^*), \nonumber \\
\h^{*}_a (\F^*) & = & \pa_{\ov{\imath}}H^*(\X,\F^*)~\k^{*i}_a (\F^*) 
+ Y^*_a (\X,\F^*), \nonumber \\
\r^{*}_a (\X^*) & = & \pa_{\ov{\hat{\imath}}}H(\F,\X^*)~\t^{*i}_a (\X^*)
- Y_a (\F,\X^*).
\label{relisom}
\eq
By differentiating the first and last equations in (\ref{relisom}) with
respect to $\X^{*j}$ and $\F^{j}$ respectively, one gets:
\bq
H_{i{\ov{\hat{\jmath}}}}~\k^{i}_{a} &=&-Y_{a\ov{\hat{\jmath}}},\nonumber\\
H_{{\ov{\hat{\imath}}}j}~\t^{*i}_{a}&=&Y_{aj},
\label{difsom}
\eq
which, in turn, allow one to write the identity:
\be
\k^{i}_{a}~Y_{bi}+\t^{*i}_{b}~Y_{a{\ov{\hat\imath}}}=0.
\label{ide}
\ee
From the algebra (\ref{lie}), and from (\ref{relisom}), we have:
\be
H_{i}~\k^{j}_{[a}\k^{i}_{b]j}+ 
H_{\ov{\hat{\imath}}}~\t^{*j}_{[a}\t^{*i}_{b]\ov{\hat{\jmath}}}=
f_{ab}^{~~c}(\h_{c}+\r^{*}_{c}),
\ee
which, by means of (\ref{ide}), can be rewriten as:
\be
\k^{j}_{[a}\h_{b]j}+ \t^{*j}_{[a}\r^{*}_{b]j}= 
f_{ab}^{~~c}(\h_{c}+\r^{*}_{c}).
\label{fun}
\ee
From holomorphicity considerations, one may set:
\bq
\k^{j}_{[a}\h_{b]j}&= & f_{ab}^{~~c} \h_{c} +ic_{ab},\nonumber\\
\t^{*j}_{[a}\r^{*}_{b]j}&= & f_{ab}^{~~c}\r^{*}_{c} -ic_{ab},
\label{kici}
\eq
where $c_{ab}=-c_{ba}$ are real constants. In the restricted 
case of a semi-simple gauge group ${\cal G}$, we may remove the $c_{ab}$'s by 
simply imposing $c_{ab}=0$ (in other cases they represent an obstruction 
to the gauging \cite{hklr}). With this restriction, one writes the
variation on the Killing potential as:   
\be
\d Y_{a}= \frac{1}{2}\l^b \biggl(\k^{i}_{[b}Y_{a]i}+\t^{*i}_{[b}Y_{a]i}
            \biggr)
        = -\l^{b}f_{ab}^{~~c} Y_{c},
\label{change}
\ee
where one has used (\ref{lie}), (\ref{relisom}) and (\ref{kici}). Now, 
from (\ref{difsom}) and (\ref{change}) we obtain the complex potential 
$Y_{a}$:
\be
Y_a = 2 f_{ab}\,^c \k^i_d\t^{*j}_c \frac{\pa^2 H}{\pa\F^{i}\pa{\X^{*j}}}
 g^{bd},
\label{y}
\ee
in which $g^{bd}$ is the inverse Killing metric.

To proceed to the covariantization of the action (\ref{action})
with respect to gauged isometries, i.e. the local version 
of the set of field transformations (\ref{global}), one introduces
a couple $(\Ld,\G)$ of real chiral and antichiral
superfield parameters respectively \cite{chv}. 
The local isometry transformations are defined as:
\be
\F' = \exp{(L_{\Ld \cdot \k})}\F,~~~~~
\X' = \exp{(L_{\G \cdot \t})}\X.
\label{FX}
\ee
The gauge sector is built up from the prepotential $V$, a real
superfield transforming such as:
\be
\exp{(L_{V^{'} \cdot \t})} = \exp{(L_{\Ld \cdot \t})}
\exp{(L_{V \cdot \t})}\exp{(-L_{\G \cdot \t})}  .
\label{V}
\ee
We modify then the action (\ref{action}) by replacing the antichiral 
superfields $(\X,~ \X^*)$ with the redefined quantities 
$({\tilde \X},~{\tilde \X^*})$ given below:
\be
\ti\X^{i} \equiv \exp{(L_{V \cdot \t})}\X^{i},~~~~~
{\tilde\X^{*i}} \equiv \exp{(L_{V \cdot \t^{*}})}{{\tilde\X^{*i}}}. 
\label{E}
\ee
Infinitesimally one has the following isometry transformation 
laws for the superfields:
\bq
&\d\F^{i}=\Ld^{a}\k^{i}_{a},~~~~~
 \d\F^{*i}=\Ld^{a}\k^{*i}_{a},&\nonumber \\
&\d{\tilde \X^i}= \Ld^{a}\t^{i}_{a},~~~~~
 \d{\tilde \X^{*i}}=\Ld^{a}\t^{*i}_{a}.&
\label{local}
\eq
It turns out moreover that the correct covariantization of (\ref{action}) 
still demands the introduction of a complex conjugated pair of
 antichiral superfields $(\ups,\ups^*)$ transforming as:
\bq
&\d\ups    = \l^{a}\r_{a}    (\X) , \nonumber \\
&\d \ups^* = \l^{a}\r^{*}_{a}(\X^*) .
\label{isoco1}
\eq
The isometry-covariant action functional is then taken to be:
\be
I_{cov} = 2\int {d^4x d^2{\q} d^2\qwt}~
\biggl [H(\F,{\tilde\X}^{*}) + 
H^{*}(\F^{*},\tilde\X) -\tilde{\ups} -
{\tilde\ups}^{*} \biggr],
\label{Sv}
\ee
which, in terms of the original variables, writes as:
\be
I_{cov} = 2 \int {d^4x d^2{\q} d^2\qwt}~\Biggl \{H(\F,\X^{*}) +
H^{*}(\F^{*},\X)  + 
2 ~Re \biggl [\frac{e^L -1}{L}~V^{a}Y^{*}_{a}(\F^*,\X)\biggr ]
\Biggr\},
\label{Sv1}
\ee
with $L\equiv L_{V.\t}$.

As mentioned in the introduction, it will be our aim hereafter to
extend the construction leading to $I_{cov}$ in (\ref{Sv1}) above
to the more general task of analyzing the gauging of N=2 
supersymmetric $\s$-model in the N=1 superspace of Atiyah-Ward. 
With this purpose in mind, one is enforced here to consider the
more restricted class of hyper-K\"ahlerian $\s$-models in
order to introduce a second set of supersymmetry field 
transformations, following in much the same way what was envisaged 
already in the last decade by Alvarez-Gaum\'e and Freedman \cite{af}.
The K\"ahlerian target space of our $\s$-model can also be taken
as a hyper-K\"ahler manifold as long as its metric tensor
$g_{{\cal I}{\cal J}}$ in (\ref{metrica}) is hermitian with 
respect to a quaternionic structure $\{J_{\cal I}^{(1)\cal J},
J_{\cal I}^{(2){\cal J}}, J_{\cal I}^{(3){\cal J}}\}$.
The tensors  $J_{\cal I}^{(x){\cal J}}$ are covariantly 
constant and generate the SU(2) algebra:
\bq
 J_{\cal I}^{(x){\cal J}} J_{\cal J}^{(y){\cal K}}=
 -\d^{xy}\d^{\cal K}_{\cal I}+\e^{xyz} J_{\cal I}^{(z){\cal K}}
\nonumber.
\eq
The complex structures are parametrized here as follows:
\be
J_{\cal I}^{(1)\cal J} 
=\left(
\begin{array}{cccc}
i\d_{i}^{j} &\0 &\0 &\0 \\
\0 & i\d_{\hat {\imath}}^{\hat {\jmath}} &\0 &\0 \\
\0 &\0 &-i\d_{\ov{\imath}}^{\ov{\jmath}} &\0 \\
\0 &\0 &\0 &-i\d_{\ov{\hat{\imath}}}^{\ov{\hat{\jmath}}} \\
\end{array}\right),
\label{comp1} 
\ee
\be
J_{\cal I}^{(2)\cal J} =
\left(
\begin{array}{cccc}
~\0~ & ~\0~ & ~\0~ & ~J_{i}^{\ov{\hat {\jmath}}}~\\
~\0~ & ~\0~ & ~J_{\hat {\imath}}^{\ov {\jmath}}~ & ~\0~\\
~\0~ & ~J_{\ov {\imath}}^{\hat {\jmath}}~&~\0~&~\0~ \\
~J_{\ov{\hat {\imath}}}^{j}~&~\0~&~\0~&~\0~\\
\end{array}\right),
\label{comp2} 
\ee
and
\be
J_{\cal I}^{(3)\cal J} =
\left(
\begin{array}{cccc}
\0 & \0 & \0 & i J_{i}^{\ov{\hat {\jmath}}} \\
\0 & \0 &  i J_{\hat {\imath}}^{\ov {\jmath}} & \0\\
\0 & -i J_{\ov {\imath}}^{\hat {\jmath}} & \0 & \0 \\
 -i J_{\ov{\hat {\imath}}}^{j} & \0 &\0 &\0 \\
\end{array}\right).
\label{comp3} 
\ee
It is the very existence of such a quaternionic structure what 
enables one to introduce a non-linearly realized supersymmetry 
in the theory. In fact, we shall see in the next section that 
the action (\ref{Sv1}) can be conveniently supplemented with 
new interaction terms which will render it invariant under N=2 
supersymmetries, while preserving its covariance under the gauged 
isometries (\ref{local}).

\section{The N=2 Supersymmetric Extension}
In this section we analyze the N=2 supersymmetric extension 
of our gauged $\s$-model in the Atiyah-Ward superspace. By following 
a reasoning similar to that of \cite{hklr}, one defines the second 
supersymmetry in terms of two sets of complex functions of the 
target coordinates, the potentials $\O^{i}\equiv \O^{i}(\F,\X^*)$ and 
$\UPS^{i}\equiv\UPS^{i}(\X,\F^*)~~(i=1,...,n)$, the field 
transformation laws being given by:
\bq
&\d \F^{i} = i{\wt{D}}^2 (\e\,\O^i),~~~~~
\d \F^{*i} =i{\wt{D}}^2 (\e\,\O^{*i}),& \nonumber \\
&\d \X^{i} =iD^2 (\z\,\UPS^i),~~~~~
\d \X^{*i} = iD^2 (\z\,\UPS^{*i}),& 
\label{ss}
\eq
where $\z$ and $\e$ are real constant chiral and antichiral scalar
superfields respectively, i.e.
\bq
 ~D_\a \e = \pa_\m \e = 0, ~~~~~ ~{\wt D_{\dot\a}} \z = \pa_\m \z = 0,  
\label{s1}
\eq
and moreover
\be
{\wt D}^2 \e = D^2 \z=0. 
\ee
The on-shell closure of the algebra of transformations in
(\ref{ss}) imposes the following constraints on the potentials:
\bq
&\O^{i},_{{\ov {\hat \jmath}}{\ov{\hat k}}}\UPS^{*j},_{n} +
\O^{i},_{{\ov {\hat \jmath}}}\UPS^{*j},_{n{\ov {\hat k}}}=0~,
~~~~~
\UPS^{i},_{{\ov \jmath}{\ov k}}\O^{*j},_{{\hat n}} + 
\UPS^{i},_{\ov \jmath}\O^{*j},_{{\hat n}{\ov k}}=0~,&
\nonumber \\
&\O^{i},_{\ov {\hat \jmath}} \UPS^{*j},_{n}= -\d^{i}_{n}~,~~~~~
\UPS^{i},_{\ov \jmath} \O^{*j},_{\hat n}=-\d^{\hat \imath}_{\hat n}~,&
\nonumber \\
&\O^{i},_{j[{\ov{\hat k}}}\O^{j},_{\ov{\hat n}]} =0~,~~~~~
\UPS^{i},_{{\hat \jmath}[{\ov k}}\UPS^{j},_{{\ov n}]}=0~,&
\nonumber \\
&{\wt{D}}^2 \O^{i}=0~,~~~~~  D^2\UPS^{i}=0~,&
\label{R1}
\eq
with the lower indices standing for derivatives with respect 
to the target coordinates. Moreover, by requiring the invariance of 
the action (\ref{action}) under (\ref{ss}) we arrive at the additional
conditions upon the functions $\O^{i}$ and $\UPS^{i}$:
\bq
&H_{i{\ov{\hat \jmath}}}\O^{i},_{\ov{\hat n}} +
 H_{i{\ov{\hat n}}}\O^{i},_{\ov{\hat \jmath}}=0,~~~~~   
 H^{*}_{{\hat \imath}{\ov \jmath}}\UPS^{i},_{\ov n} +
 H^{*}_{{\hat \imath}{\ov n}}\UPS^{i},_{{\ov \jmath}} =0, &
\nonumber\\
&H_{i{\ov{\hat n}}}\O^{i},_{{\ov{\hat \jmath}}{\ov{\hat k}}}+
 H_{i{\ov{\hat \jmath}}{\ov{\hat k}}}\O^{i},_{\ov{\hat n}}=0,~~~~~
 H^{*}_{{\hat \imath}{\ov n}}\UPS^{i},_{{\ov \jmath}{\ov k}}+
 H^{*}_{{\hat \imath}{\ov \jmath}{\ov k}}\UPS^{i},_{\ov n}   =0, & 
\nonumber\\
&H_{i{\ov{\hat \jmath}}}\O^{i},_{{\ov{\hat n}}k}+
 H_{i{\ov{\hat \jmath}}k}\O^{i},_{\ov{\hat n}} =0,~~~~~
 H^{*}_{{\hat \imath}{\ov \jmath}} \UPS^{i},_{{\ov n}{\hat k}}+
 H^{*}_{{\hat \imath}{\ov \jmath}{\hat k}}\UPS^{i},_{\ov n} =0, &   
\label{R2}
\eq
together with their complex conjugated counterparts. At this
point, by means of a careful inspection of eqs. (\ref{R1})
and (\ref{R2}), one observes that the functions $\O^{i}$
and $\UPS^{i}$ are encompassing in their structure all the
important features of the hyper-K\"ahlerian geometry \cite{hklr,hiklr}.
Indeed, this property can be made even more apparent if we 
introduce the identifications:
\be
 J_{\ov{\hat \imath}}^{~j}= \O^{j},_{\ov{\hat \imath}}~,~~~~~
 J_{\ov \imath}^{~\hat \jmath}=\UPS^{j},_{\ov \imath}~,~~~~~
 J_{{\hat \imath}}^{~\ov \jmath}= \O^{*j},_{\hat \imath}~,~~~~~
 J_{i}^{~\ov{\hat \jmath}}=\UPS^{*j},_{i}~,  
 \label{s}
\ee
in the complex structures (\ref{comp2}) and (\ref{comp3}).

Furthermore, from the assumption of triholomorphicity of the Killing 
vectors with respect to the quaternionic structure,
one can define the potentials $P_{a}^{(+)}\equiv P_{a}^{(+)}(\F,\X)$
and $P_{a}^{(-)}\equiv P_{a}^{(-)}(\F^*,\X^*)$ such that 
$P_{a}^{(-)}=(P_{a}^{(+)})^*$
and
\bq
&k_{a}^{i}\o^{(+)}_{ij} = -P_{a}^{(+)},_{j}~,~~~~~
k_{a}^{*i}\o^{(-)}_{{\ov \imath}{\ov \jmath}} = 
-P_{a}^{(-)},_{\ov \jmath}~,&
\label{P1}\\ 
&\t_{a}^{\hat \imath}\o^{(+)}_{{\hat \imath}{\hat \jmath}} = 
-P_{a}^{(+)},_{\hat \jmath}~,~~~~~
\t_{a}^{*\hat \imath}\o^{(-)}_{{\ov{\hat \imath}}{\ov{\hat \jmath}}} =
 -P_{a}^{(-)},_{{\ov{\hat \jmath}}}~,
\label{P2}
\eq
with
\bq
\o^{(+)}_{ij}=-2H_{j{\ov{\hat k}}}\UPS^{*k},_{i}~,~~
\o^{(-)}_{{\ov \imath}{\ov \jmath}}=
 -2H^{*}_{{\ov \jmath}{\hat k}}\UPS^{k},_{\hat \imath}~,~~
\o^{(+)}_{{\hat \imath}{\hat \jmath}}=
 -2H^{*}_{{\hat \jmath}{\ov k}}\O^{*k},_{\hat \imath}~,~~
\o^{(-)}_{{\ov{\hat \imath}}{\ov{\hat \jmath}}}=
 -2H_{{\ov{\hat \jmath}}k}\O^{k},_{\ov{\hat \imath}}. 
\label{O1}
\eq
From eqs. (\ref{relisom}) above and from the formulae expressing the 
chiral and antichiral functions (\ref{P1},\ref{P2}) we derive some 
useful relations involving the Killing potentials $Y_{a}(\F,\X^*)$:
\bq
P^{(+)}_{a,j}\O^{j}_,{\ov {\hat \imath}}= -2Y_{a,{\ov {\hat \imath}}}
&\rightleftharpoons & P^{(+)}_{a,i}= 2 Y_{a,{\ov {\hat\jmath}}}
\UPS^{*j},_{i}
\nonumber \\
P^{(+)}_{a,{\hat \jmath}}\UPS^{j},_{\ov \imath}= 2Y^{*}_{a,{\ov\imath}}
&\rightleftharpoons &P^{(+)}_{a,{\hat \imath}}= 
-2 Y^{*}_{a,{\ov {\hat \jmath}}}\O^{*j},_{{\hat \imath}} 
\nonumber \\
P^{(-)}_{a,{\ov \jmath}}\O^{*j},_{\hat \imath}=-2Y^{*}_{a,{\hat\imath}}
&\rightleftharpoons & P^{(-)}_{a,{\ov \imath}} = 
2Y^{*}_{a,{\hat \jmath}}\UPS^{j},_{\ov \imath}
\nonumber \\
P^{(-)}_{a,{{\hat {\ov \jmath}}}} \UPS^{*j},_{i}= 
2Y_{a,i} 
&\rightleftharpoons & P^{(-)}_{a,{\ov {\hat \imath}}}=
-2Y_{a,j}\O^{j},_{{\ov {\hat \imath}}}. 
\label{ahhh}
\eq
To obtain $P^{(+)}_{a}$ and $P^{(-)}_{a}$ one observes that the complex
functions
\be
U_{a}= P^{(-)}_{a} - P^{(+)}_{a} -2iY_{a} +2iY^*_{a}
\label{U}
\ee
do satisfy the following differential equations:
\bq
 (\pa_{i} + i\UPS^{*j}_{~~,i}\pa_{\ov{\hat \jmath}})~U_{a}=0, 
\label{U1} \\
 (\pa_{\hat \imath} +i\O^{*j}_{~~,\hat\imath} \pa_{\ov\jmath})~U_{a}=0, 
\label{U2}\eq
the complex conjugated, $U^{*}_{a}$, obeying the complexified analogs 
thereof. Actually, eqs. (\ref{U1}), (\ref{U2}) are specifying the 
$U_{a}$'s (resp. $U^{*}_{a}$'s) as holomorphic functions (resp. 
antiholomorphic functions) relatively to a non-canonical complex 
structure \cite{hklr}. From the definition given in (\ref{U}), one 
can write:
\bq
&U + U^* = 4(-iY + iY^*) 
\label{c0}
\eq
and
\bq
&U -U^*= 2(P^{(-)}-P^{(+)}). 
\label{c}
\eq
Now, from the holomorphicity and the gauge transformation
of $Y_{a}$:
\be
\d Y_{a}=-\l^{b}f_{ab}\,^{c} ~ Y_{c},
\label{y1}
\ee
one arrives at
\bq
\d P^{(+)}_{a} &=& -\l^{b} f_{ab}\,^{c} ~ P^{(+)}_{c},
\label{dp1}
\eq
and
\bq
\d P^{(-)}_{a} &=& -\l^{b} f_{ab}\,^{c} ~ P^{(-)}_{c},
\label{dp2}
\eq
where the gauge group was assumed to be semi-simple, which implies the 
absence of obstructions in (\ref{y1}), (\ref{dp1}) and (\ref{dp2}) above. 
On the other hand we have:
\be
 \d P^{(+)}_{a} = \l^{b} \biggl( k^{i}_{b} P^{(+)}_{a},_{i} + 
                  \t^{i}_{b} P^{(+)}_{a},_{\hat \imath} \biggr),
\label{dp3}
\ee
which, by comparison with (\ref{dp1}) and use of the first equations of 
(\ref{P1}) and (\ref{P2}), gives us the following:
\be
 P^{(+)}_{a}=f_{a}\,^{bc}\biggl( k^{i}_{c}k^{j}_{b}\o^{(+)}_{ji} +
 \t^{i}_{c}\t^{j}_{b}\o^{(+)}_{{\hat \jmath}{\hat \imath}} 
 \biggr).
\label{PP}
\ee
Through complex conjugation, one also has:
\be
P^{(-)}_{a}=
 f_{a}\,^{bc} ~ 
 \biggl( k^{*i}_{c}k^{*j}_{b}\o^{(-)}_{{\ov \jmath}{\ov \imath}} 
 + \t^{*i}_{c}\t^{*j}_{b}\o^{(-)}_{{\ov{\hat \jmath}}
 {\ov{\hat \imath}}} \biggr).
\label{PP*}
\ee

We now turn to the construction of the N=2 supersymmetric gauge sector 
in the N=1 superspace of Atiyah-Ward. In \cite{ggrs}, Gates et al.
succeeded in writting down a set of non-linear supersymmetry
transformations for a certain N=2 gauge-supermultiplet in N=1 Minkowski 
superspace. We adopt a similar approach here: our K\"ahlerian gauge 
supermultiplet consists of a chiral scalar superfield $S$ and an 
antichiral scalar superfield $T$, together with the vector superfield 
$V$ of the previous section. All the three superfields are real and take 
values in the adjoint representation of the isometry gauge group 
${\cal G}$. We propose the non-linear supersymmetry transformations 
on gauge superfields 
as below:
\bq
\d S &=& i W^\a D_\a \z,
\nonumber \\
\d T &=& i \tilde W^{\dot\a}{\wt{D}_{\dot\a}}\e, 
\nonumber \\
 e^{-iV} \d e^{iV}&=& \e~e^{-iV}S e^{iV} - \z~T,
\label{ss2}
\eq
where the real scalar superfield parameters $(\e,\z)$ are the ones
appearing in the supersymmetry transformations (\ref{ss}) for the
matter sector; the gauge superfield-strengths are defined to be:
\bq
W_{\a}&\equiv & i{\wt{D}}^{2}\biggl( e^{iV} D_{\a} e^{-iV} \biggr),
\nonumber \\
{\tilde W_{\ad}}&\equiv & i D^{2}\biggl( e^{-iV} {\wt{D}_{\ad}} 
e^{iV} \biggr),
\label{c2}
\eq
they are covariant under gauge transformations of the type
\bq
e^{-iV}\d_{g} e^{iV}&=&i\biggl( e^{-iV}\Ld e^{iV}-\G \biggr).
\eq
One has also to consider gauge transformation laws for the scalar 
gauge superfields:
\bq
\d_{g}S = i[\Ld,S],~~~~~
\d_{g}T = i[\G,T].
\eq
At this stage, we are ready to present the fully gauged N=2 
supersymmetric non-linear $\s$-model in terms of N=1 superfields 
of the Atiyah-Ward superspace. As stated previously, this task is 
accomplished by suplementing the action (\ref{Sv1}) with new interaction 
pieces such as to render the second supersymmetry, i.e. (\ref{ss}) 
and (\ref{ss2}), a further invariance of the model \cite{hklr}. Our main 
result is:
\bq
&I_{cov} =  2 \int  {d^4xd^2{\q}d^2\qwt}~ \biggl\{ H(\F,\X^{*}) 
            +  H^{*}(\F^{*},\X)  + 
            2~Re \biggl [\frac{e^{\hat{\cal L}} -1}{\hat{\cal L}}~V^{a}
            Y^{*}_{a}(\F^*,\X)\biggr] 
            -\frac{1}{2} S^{a}{\tilde T}_{a} \biggr\} + 
\nonumber \\
&     - \frac{1}{16}\int{d^4xd^2{\q}}~\biggl\{g_{ab}W^{a\a}W^{b}_{\a}
      - 4i S^{a}\biggl[ F_{a}(\F)+F_{a}^*(\F^*)\biggr]\biggr\} +
\nonumber \\
&     - \frac{1}{16}\int{d^4xd^2\qwt}~\biggl\{g_{ab}{\tilde W^{a\dot\a}}
         {\tilde W^{b}_{\dot\a}} 
      - 4i T^{a} \biggl[ G_{a}(\X) + G_{a}^*(\X^*)\biggr]\biggr\},
\label{scov}
\eq
where we have made implicit use of the splittings in the functions 
$P_{a}^{(+)}$ and $P_{a}^{(-)}$ in (\ref{PP}) and (\ref{PP*}):
\be
P^{(+)}_{a}=F_{a}(\F)+G_{a}(\X),~~~~~
P^{(-)}_{a}=F^{*}_{a}(\F^*)+G^{*}_{a}(\X^*).
\ee
Finally, it is straightforward to check the invariance of (\ref{scov}) 
under (\ref{ss2}) and the (gauge covariant) supersymmetry transformations 
for the matter superfields:
\bq
&\d \F^{i} =i{\wt{D}}^2 
 \biggl(\e~\O^i(\F,e^{2{\hat{\cal L}}^{*}}\X^*)\biggr),
 ~~~~~
 \d \F^{*i}=i{\wt{D}}^2 
 \biggl(\e~\O^{*i}(\F^*,e^{2{\hat {\cal L}}}\X)\biggr),&
 \nonumber \\
&\d \X^{i} =iD^2 \biggl(\z~\UPS^i(e^{-2{\cal{L}}^{*}}\F^*,\X)\biggr),
 ~~~~~
 \d \X^{*i} =iD^2 \biggl(\z~\UPS^{*i}(e^{-2\cal{L}}\F,\X^*)\biggr),&
\eq
in which
\be
{\cal L} = V^{a} k_{a}^{i}\frac{\pa}{\pa\F^i},~~~~~
{\hat{\cal L}} = V^{a}\t_{a}^{i}\frac{\pa}{\pa\X^i}.
\ee
Indeed, one may impose the Wess-Zumino gauge condition, i.e. $V^{3}=0$, 
and to verify the invariance of $I_{cov}$ under the non-linear 
supersymmetry transformations at each order in the prepotential $V$. 
It should be observed once more that due to the presence of some 
gauge-algebraic obstructions \cite{hklr,whr}, the supersymmetric gauging 
expressed in (\ref{scov}) will only hold for semisimple gauge groups 
${\cal G}$, in which case one can always determine the potentials 
(\ref{y}), (\ref{PP}) and (\ref{PP*}). 

\section{Concluding Remarks}

We have explicitly constructed a class of N=2 supersymmetric non-linear
$\s$-models coupled to a super-Yang-Mills gauge sector in the N=1
superspace of Atiyah-Ward. In order to perform this gauge coupling,
one makes use of a general formalism introduced by Hull et al. in 
\cite{hklr}, gauging the isometries of the associated (hyper-K\"ahler)
target manifold. We observe then that, also in the Atiyah-Ward 
superspace, it is possible to obtain the specific potentials needed 
for the referred gauging of the hyper-K\"ahlerian $\s$-model, namely 
the Killing potential (\ref{y}) (which is complex here) and the 
so-called momentum maps (\ref{PP}) and (\ref{PP*}).

The gauge-invariant supersymmetric $\s$-model obtained in the previous
section may have some interesting applications in connection with the
study of gauge dynamics of supersymmetric gauge theories in lower 
dimensions. In fact, by suppressing one time coordinate in the action
(\ref{scov}), one may in principle arrive at new supersymmetric field
models in three Minkowskian dimensions. The latter type of theories 
could then be regarded as an alternative scenario for checking the 
consequences of the duality hypothesis of four dimensions, following 
in much the same way what has been proposed in the recent literature 
\cite{sw}.

\section*{Acknowledgements}

The two authors express their gratitude to Jos\'e A. Helay\"el-Neto 
for many interesting remarks and insightful discussions, and to 
Hassan Zerrouki for pointing out some references. They also thank the 
Brazilian agencies C.A.P.E.S. and C.N.Pq. for the financial support.

\end{document}